\documentclass[aps,twocolumn,prb,floatfix,showpacs,preprintnumbers,superscriptaddress,amsmath,amssymb]{revtex4-2}

\usepackage{mathtools}
\usepackage{bbold}
\usepackage{amsmath}
\usepackage{multirow}
\usepackage{amsfonts}
\usepackage{bm}
\usepackage{amssymb}
\usepackage{mathrsfs}
\usepackage{hyperref}
\usepackage[pdftex]{graphicx}

\begin{document}
\title{Optical chromatography ultra small particles by Brownian motion in tilted optical potential induced by bound states in the continuum}
\author{Evgeny N. Bulgakov}
\affiliation{Kirensky Institute of Physics, Federal Research Center KSC SB
RAS, 660036, Krasnoyarsk, Russia}
\author{Galina V. Shadrina}
\affiliation{Institute of Computational Modelling SB RAS, Krasnoyarsk, 660036, Russia}
\thanks{galiysha1996@gmail.com}
\date{\today}
\begin{abstract}
We investigate sorting Rayleigh optical particles up to several nanometers in size during Brownian motion in an tilted periodic potential with multiple deep wells. The wells are induced which by optical bound states in the continuum in a system of parallel photonic crystal slabs immersed in a liquid. The Brownian dynamics of the particles is significantly altered by resonant optical forces leading to the complete spatial separation of particles with a size difference of approximately $1\%$ during the diffusion process. In addition, the possibility of creating an integrated platform for continuous optical sorting is discussed.
\end{abstract}

\maketitle

\section{Introduction}
The manipulation of micrometer and submicrometer particles using noncontact forces is an important area of contemporary applied research~\cite{Jon2008,Gao2017}.
Particularly, in the last two decades it has been proposed to use various optical beams and nanosized devices for mechanical action on nanoparticles in order to control their position and motion with high precision. For this purpose beams that are both linearly polarized or have angular momentum are suitable, making it possible to generate optical forces and torques for optical coding, trapping, binding, sorting, and moving nanoparticles from one place to another. As a result optical manipulation has become an effective technique in various scientific areas, such as optics, atomic physics, biological sciences, and chemistry.

Optical forces are most frequently used  for particle trapping. In optical tweezers the optical beam is focused through a lens for increasing the EM field intensity and, consequently, the optical forces inside the trap. However, this enhancement is insufficient for Rayleigh particles since the optical forces for a particle with a micrometer size drop cubically with the particle's radius. 
An alternative approach to enhance optical forces utilizes the near-field in the immediate vicinity of the optical microstructure with a high-Q factor resonant mode excited by an external field. 
In this situation the gradient optical forces are amplified by roughly a factor of $Q$ and can surpass the Brownian forces~\cite{Mao2022,Hu2018,Zaman2019,Erickson2011,Davis2007, Deych2023,Bulgakov2023,Bulgakov2020,Wang22,Yang2023}.

The standard method to achieve resonant forces with an extraordinarily high $Q$-factor is application of optical bound states in the continuum (BIC) \cite{Koshelev2019}. The ideal BICs can exist exclusively within infinite systems. However, in a finite system quasi-BIC modes occur retaining the features of the ideal BIC but possessing a finite $Q$-factor that scales with the relation $Q \sim N^{\alpha}$, where $N$ represents the number of elementary cells. An example of this is a linear array of dielectric resonators, where $\alpha$ can be 2 or 3. If a further increase of the $Q$ factor is require, on can use super-BIC modes \cite{Hwang2021,Bulgakov2023b}. In \cite{Hwang2021}, a super-BIC mode with $Q \approx 10^8$ was found for the case of $N=50$ cells. Thus, it is possible to achieve an enhancement of the EM field in the near-field by many orders of magnitude if quasi-BIC resonant modes are used for this purpose.

For optical sorting specially engineered profiles of optical intensity are used. The optical forces in such profiles exhibit a complicated dependence on the shape and size of particles as well as on the refractive index, thus, providing a basis for effective sorting. Among the sorting methods, one can distinguish static sorting, which does not use liquid flow \cite{imr2006,Brzobohat2013,Jkl2008,Jkl2014} and dynamic sorting in which the host fluid is in motion \cite{Bulgakov2023,Ladavac2004,Zhao2021} 
For nanometer-sized particles the motion within an optical trap becomes complex due to the influence of random Brownian forces.
The optical potential significantly changes Brownian dynamics and generally the motion has to be described using the Langevin or the Fokker-Planck (FP) equations. The latter describes the statistical behavior of the diffusing particle via a distribution function dependent on both spatial coordinates and time, the external forces being directly into the equations \cite{Davis2007, Zaman2019, Grima2007, Reimann2002, Wu2016,Evstigneev2008, Harada1998, Han2006, Risken1984}. The extreme situation of deep potential wells compared to $kT$ is well described by Kramers theory. For example, \cite{McCann1999} presented results on the thermally induced transitions of a Brownian particle between adjacent optical traps. The rate at which transitions occur aligns very well with the predictions made by the Kramers formula~\cite{Kramers1940}.

Even more complex Brownian dynamics is observed if the potential relief consists of many potential wells similar to a tilted periodic potential (washboard potential) \cite{Evstigneev2008,Paterson2005,Dholakia2007,Lindenberg2007,Gleeson2006}.
This tilted potential can be created using several rotating optical traps \cite{Evstigneev2008}, or using a Bessel beam \cite{Paterson2005,Dholakia2007}. As it turns out, the transport properties of particles in such a potential strongly depend on the physical characteristics of the particles such as size and permittivity, and for this reason such potentials can be used effectively for the spatial separation of particles \cite{Bulgakov2024w,Paterson2005,Dholakia2007,Lindenberg2007,Gleeson2006}.

The aim of this work is to study the sorting of Rayleigh-Brownian optical particles of the order of a few nanometers in size using a tilted periodic potential generated by excitation of a resonant optical mode with a high $Q$-factor in a system of two parallel photonic crystal (PhC) slabs. The best candidate for such a mode is an optical bound state in the continuum, since it formally has an  infinite $Q$-factor and is easily excited by an external source resulting in a field enhancement between the PhC slabs \cite{Marinica2008}, and formation of the desired periodic potential that tightly localizes even ultrasmall nanoparticles.

Despite the rigid localization, due to Brownian fluctuations the particles are able to hop from one well to another well changing their position. In a long time interval, a particle travels along the periodic potential with an average velocity $v_{\text{drift}}$. At the same time the interwell diffusion with a diffusion coefficient $D$ leads to spreading of the initially localized distribution function. The depth of the optical potential for a Rayleigh spherical particle is $\sim r^3$ with $r$ being the size of the particle. A slight change in the particle radius for a deep potential leads to a significant change in both the drift velocity and the diffusion coefficient. These changes are sufficient for particles of similar sizes to be spatially separated over a spatial interval of $\sim 100$ potential wells after some time, despite the fact that at the initial moment in time they occupied at the same position. 

In our study, we propose two options for creating a tilted periodic potential. One option is to maintain constant liquid flow between the long PhC slabs. Another option is to employ finite PhC slabs without fluid motion. In the case of finite slabs containing $N$ periods, the external source excites a quasi-BIC with envelope profile that resembles a standing wave, so that the depth of the wells increases monotonically from the periphery to the center of the structure.

In addition, the article discusses a method for removing spatially separated particles from the system. For this an integrated platform can be created for continuous optical sorting. In this set-up, the input is a mixed constant flow of particles of different sizes, and the outputs are  constant flows of sorted particles carried away through different spatially separated channels.

\section{Fokker-Planck diffusion equation and transition to a tight-binding model}

\begin{figure}[h!]
\centering
\includegraphics[width=8.5cm]{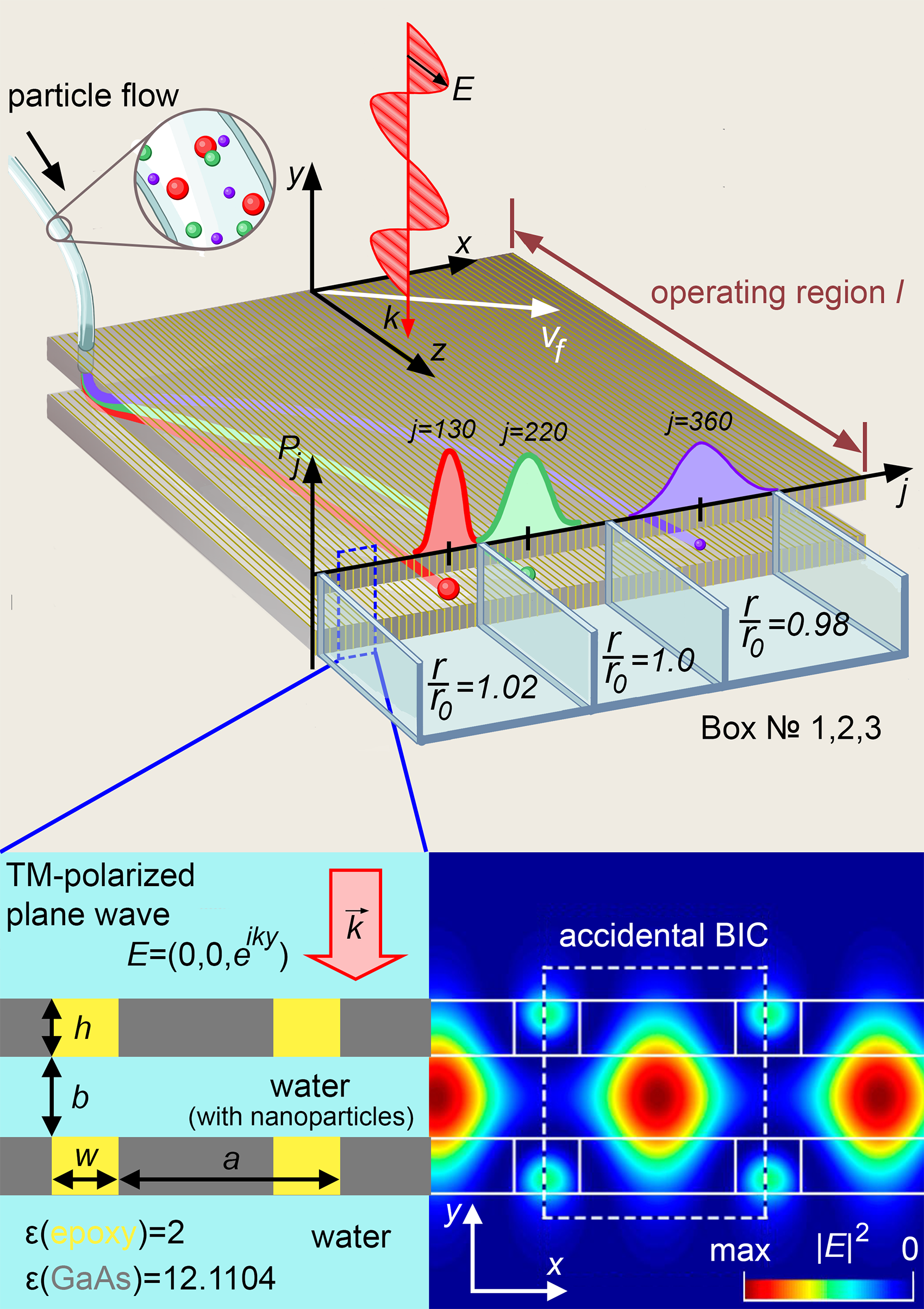}
	\caption{ Photonic crystal structure of two dielectric slabs with periodically modulated refractive index. The incident $E_z$ polarized plane wave excites the resonant optical mode, the profile of which is illustrated in the inset. The motion of liquid causes nanoparticles to diffuse in a periodic potential, resulting in the spatial distribution of particles based on their sizes.}
 \label{ris:fig1}
\end{figure}

The photonic dielectric structure is a double grating in the form of two parallel slabs immersed in water, as shown in Fig.\ref{ris:fig1}.

\begin{figure}[h!]
\centering
\includegraphics[width=8.5cm]{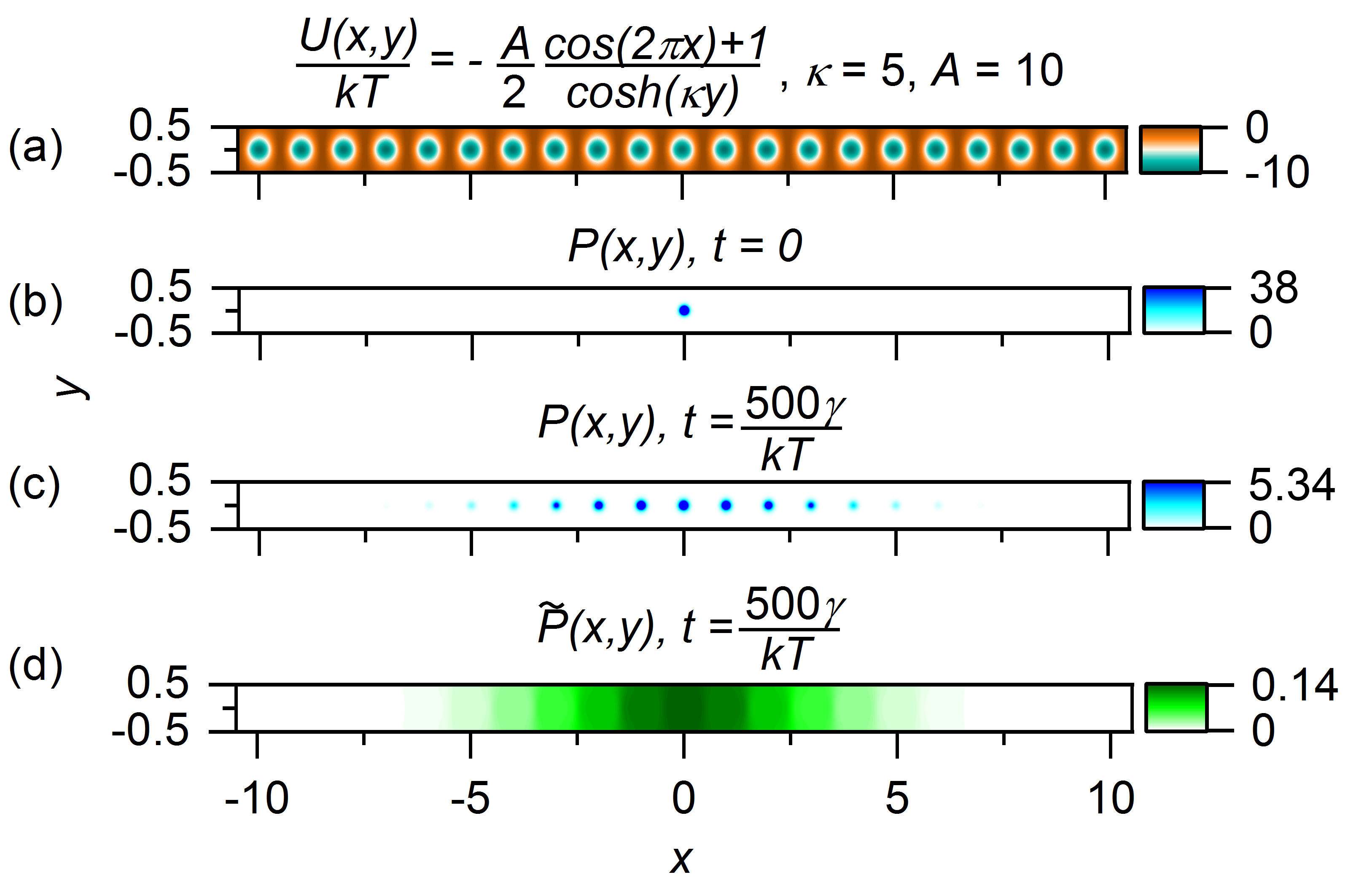}
	\caption{\label{ris:fig2}  Solution of the FP equation in the case of the model potential \eqref{three}, (a) model potential \eqref{three}, (b) probability density $P({\bf r},t)$ at the initial time $t=0$, (c) probability density at the time $t=500\gamma/kT$, (d) function $\tilde{P}({\bf r},t)$ at the time $t=500\gamma/kT$.}
\end{figure}

The optical potential is excited in the gap between the slabs, where the mixed particles are injected. The optical potential $U_0(x,y)$ does not depend on the coordinate $z$ due to the shape of the excited TM mode. Here we shall not discuss the physical reason for the appearance of $U_0(x,y)$ and its specific form. The main feature that is essential is that it is periodic in the $x$ direction $U_0(x+a,y)=U_0(x,y)$, and second, it is sufficiently deep $\Delta U \gg k_BT$. For the sorting process, it is important that the potential is tilted. For this purpose, a constant external mechanical force $F$ is added to $U_0(x,y)$. In our case, this force is the friction force in the fluid flow that moves with a constant velocity $\bf{v}_f$ relative to the slabs. Since the motion in the $(x,y)$ plane is independent of the motion along $z$ (the variables are separated), then the probability density function $P({\bf r},t)$, ${\bf r}=(x,y)$ in the overdamped regime satisfies the two-dimensional FP equation 
\begin{equation}\label{one}
\frac{\partial P({\bf r},t)}{\partial t} = \hat{L} P({\bf r},t) \mbox{ ,}
\end{equation}
here operator $\hat{L}$ reads as
\begin{equation}\label{two}
\hat{L} = \frac{1}{\gamma} \sum_{j=1,2} \frac{\partial}{\partial x_j} \bigg( k_BT \frac{\partial}{\partial x_j} + \frac{\partial U({\bf r})}{\partial x_j} \bigg) \mbox{ ,}
\end{equation}
\begin{center}
$U({\bf r}) = U_0({\bf r}) - Fx  \mbox{ ,}$
\end{center}
here $\gamma$ is friction coefficient~\cite{Risken1984}. We assume the potential $U_0(\bar{r})$ as shown in Fig.\ref{ris:fig2}a in analytical form 
\begin{equation}\label{three}
\begin{split}
\frac{U_0({\bf r})}{k_BT} = - \frac{A}{2} \frac{(cos(2 \pi x) + 1)}{cosh(\varkappa y)} \mbox{ ,}
\\
A=10, \varkappa=5 \mbox{ .}
\end{split}
\end{equation}

Equation~\eqref{one} is solved in the region $|y|<1/2$ with the boundary conditions of the absence of probability current through the lower $y=-1/2$ and upper $y=+1/2$ boundaries,
\begin{equation}\label{four}
-\frac{1}{\gamma} \bigg( k_BT \frac{\partial}{\partial y} + \frac{\partial U}{\partial y} \bigg) P({\bf r},t) \bigg|_{y=\pm 1/2} = 0 \mbox{ .}
\end{equation}
Solving equation \eqref{one} with boundary conditions \eqref{four} on a long time interval is a difficult problem, however, in our quasi-one-dimensional case, when diffusion occurs mainly along the $x$ axis, an obvious simplification is possible, leading to the one-dimensional FP equation. 
%
Figure~\ref{ris:fig2}(c) shows the numerical solution of equation \eqref{one} for the probability density after a time interval $t=500\gamma/(kT)$, if at the initial moment only the central well is populated (Fig.\ref{ris:fig2}(b)). The probability density is distributed strictly over the centers of the wells, where the potential energy is minimal. In addition, thermodynamic equilibrium along the $y$ coordinate is actually already established. To verify this, let us consider the function $\tilde{P}({\bf r},t) = e^{U({\bf r})/k_BT} P({\bf r},t)$. Its form is shown in the figure \ref{ris:fig2}(d). The function $\tilde{P}({\bf r},t)$ is practically independent of the coordinate $y$, but depends on $x$. $\tilde{P}({\bf r},t)$ becomes a constant only upon reaching complete thermodynamic equilibrium, which is impossible for an infinite system. This gives us a reason to switch in the FP equation from $P({\bf r},t)$, to a new function $\tilde{P}({\bf r},t)$, and then average the equation (\ref{one}) over the coordinate $y$. The equation for $\tilde{P}({\bf r},t)$ is
\begin{equation}\label{five}
\gamma e^{-\frac{U}{kT}} \frac{\partial \tilde{P}}{\partial t} = kT \sum_{i=1,2} \frac{\partial}{\partial x_i} (e^{-U/kT} \frac{\partial}{\partial x_i} \tilde{P}(\bar{r},t))   \mbox{ .}
\end{equation}
with boundary conditions: $\frac{\partial }{\partial y} \tilde{P}({\bf r},t) \big|_{y=\pm 1/2} = 0$. Averaging the left and right hand parts (\ref{five}) over the $y$ coordinate  
$\int^{1/2}_{-1/2}(...) dy$ we obtain 
\begin{multline}\label{six}
\gamma \frac{\partial }{\partial t} \int^{1/2}_{-1/2} e^{-U({\bf r})/kT} \tilde{P}({\bf r},t) dy= \\ 
=kT \frac{\partial }{\partial x}  \int^{1/2}_{-1/2} e^{-U({\bf r})/kT} \frac{\partial \tilde{P}({\bf r},t)}{\partial x} dy \mbox{ .}
\end{multline}
The term containing differentiation with respect to $y$ in equation \eqref{six} disappears due to the boundary conditions. As we have established, the function $\tilde{P}({\bf r},t)$ is practically independent of the coordinate $y$ and therefore the derivative $\frac{\partial \tilde{P}({\bf r},t)}{\partial x}$ on the right-hand side \eqref{six} can be taken out from the integral. If we now define the effective one-dimensional potential, according to the formula
\begin{equation}\label{seven}
 e^{-U_{eff}(x)/kT} = \int^{1/2}_{-1/2} e^{-U({\bf r})/kT} dy          \mbox{ ,}
\end{equation}
then for the function $\tilde{P}(x,t)$ we arrive at a one-dimensional equation
\begin{equation}\label{eight}
\gamma e^{-U_{eff}(x)/kT} \frac{\partial \tilde{P}(x,t)}{\partial t} 
=kT  \frac{\partial}{\partial x} \bigg(  (e^{-U_{eff}(x)/kT} \frac{\partial \tilde{P}(x,t)}{\partial x} \bigg)   \mbox{ .}
\end{equation}
Accordingly for the one-dimensional probability density $W(x,t) = e^{-U_{eff}(x)/kT} \tilde{P}(x,t)$ we obtain the one-dimensional FP equation
\begin{equation}\label{nine}
\gamma  \frac{\partial W(x,t)}{\partial t} = \frac{\partial}{\partial x} \bigg( k_BT \frac{\partial}{\partial x} + \frac{\partial U_{eff}(x)}{\partial x}  \bigg) W (x,t)  \mbox{ .}
\end{equation}
Thus, long time diffusion when the motion is essentially limited along the $y$ coordinate, can be described by a one-dimensional FP (\ref{nine}) with an effective potential. 

\begin{figure}[h!]
\centering
\includegraphics[width=8.5cm]{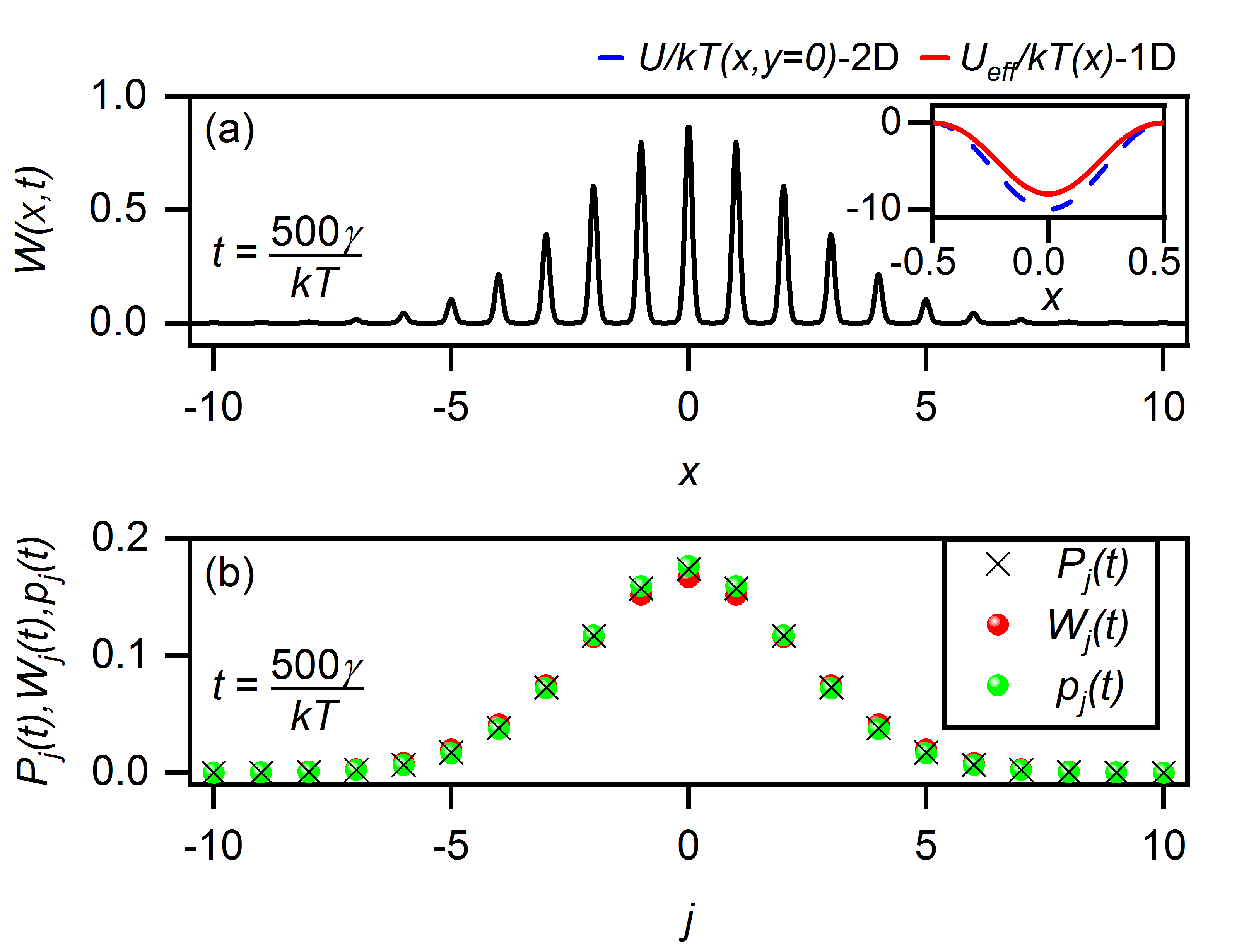}
	\caption{\label{ris:fig3} Modeling of 2D diffusion using effective one-dimensional potential $U_{eff}(x)$. (a) inset, Model potential $U(x, y=0)$~--- dash blue line and $U_{\text{eff}}(x)$~--- red line, (a) solution of equation (\ref{nine}) at time $t=500\gamma/kT$ (b) probability of being in the $j$-th potential well calculated in three ways for time $t=500\gamma/kT$, crosses~--- solution of 2D FP equation \eqref{one}, red circles~--- 1D FP equation \eqref{nine}, green circles~--- solution of discrete model.}
\end{figure}

As an example, in figure \ref{ris:fig3}(a)(inset) we show the calculated effective potential profile $U_{eff}(x)/kT$ for an essentially two-dimensional potential shown in Fig.~\ref{ris:fig2}(a). 
The solution of equation \eqref{nine} with $U_{eff}(x)$ found by formula \eqref{seven} is shown in figure \ref{ris:fig3}(a). To compare the results obtained in different ways, we will calculate the probability of finding a particle in the $j$-th potential well $W_j(t) = \int^{j+1/2}_{j-1/2} W(x,t) dx $ in the case of the one-dimensional model and $P_j(t) = \int^{j+1/2}_{j-1/2} \int^{1/2}_{-1/2} P({\bf r},t) dy dx $ in the case of the two-dimensional model. Figure \ref{ris:fig3}(b) shows the probabilities $W_j, P_j$ for the time $t=500\gamma/kT$ with the initial conditions being the same. The agreement between the results is quite convincing. If, however, when solving equation \eqref{nine} we take $U(x,y=0)$ as one dimension potential (a natural choice) then there is no agreement between the one-dimension and the two-dimension simulations.

One can go further in simplifying the original $2D$ problem \eqref{one} with potential wells. When the potential $U_{\text{eff}}(x)$ is a set of potential wells that are deep compared to $k_BT$, it is intuitively clear that the continuous diffusion equation \eqref{nine} can be approximated by a simple discrete (tight-binding) model with some hopping probabilities between adjacent wells. Such a discrete model can be derived directly from equation \eqref{nine} \cite{Nguyen2016,Challis2018,Challis2013}. In essence, this method is quite analogous to the tight-binding method, which originates from the quantum mechanics of electrons in solids~\cite{Kittel2005}. 

Let $p_j(t)$ be the probability that a particle is in the $j$-th well. Then the probabilities satisfy the system of equations
\begin{equation}\label{ten}
\begin{split}
\frac{\partial p_j(t)}{\partial t} = K_{j,j-1}p_{j-1}(t) + K_{j,j+1}p_{j+1}(t)-\\
- (K_{j+1,j} + K_{j-1,j}) p_{j}(t)               \mbox{ ,}    
\end{split}
\end{equation}
here $K_{j,j'}$ are the rates of hopping between the adjacent wells. 
A method for finding matrix elements $K_{j,j'}$ via a biorthogonal basis of a nonselfadjoint evolution operator (the right-hand side of the equation \eqref{nine}) is given in ~\cite{Nguyen2016}. 
Our numerical simulation of diffusion is greatly simplified by using the discrete equation \eqref{ten}. Application of this discrete model to a wide range of one-dimensional potentials \cite{Nguyen2016} shows that the dynamics of the system at long times is well approximated if the depth of the potential wells is greater than $5k_BT$.

The results of the calculations of $p_j(t)$ from equation \eqref{nine} using the transition to the discrete model \eqref{ten} are shown in the figure \ref{ris:fig3}(b) by green circles. In the case of a tilted periodic potential, there are only two matrix elements of the jump $K_+$~--- forward, $K_-$~--- backward. The average drift velocity of a particle along the $x$ axis
\begin{equation}\label{ten1}
v = a(K_+ - K_-)               \mbox{ ,}
\end{equation}
and the diffusion coefficient
\begin{equation}\label{ten2}
D = \frac{a^2}{2}(K_+ + K_-)               \mbox{ ,}
\end{equation}
is expressed through matrix elements, where $a$ is the period of the potential. The values of $K_{\pm}$ can be found approximately using analytical formulas
\begin{equation}\label{ten3}
K_{\pm} = K_0 exp \bigg(\pm \frac{1}{2}\frac{Fa}{kT} \bigg)               \mbox{ ,}
\end{equation}
\begin{equation*}
K_0 = \frac{\sqrt{V''(x_{min}) V''(x_{max})}}{2 \pi \gamma} exp \bigg(-\frac{\Delta V}{kT} \bigg) \mbox{ ,}
\end{equation*}
here $K_0$ is the Kramers escape rate, $V''(x_{min})$, $V''(x_{max})$ is the curvature of periodic potential at minimum and maximum points, $\Delta V$ is the potential well depth. 

In order to understand the principle of sorting particles by radius, let us consider a model potential that includes a dependence on the particle radius $r$, 
\begin{equation}\label{ten4}
\frac{U(x)}{k_BT} = r^3 V_0(x) - e_0 r x            \mbox{ .}
\end{equation}
Next we consider dimensionless coordinate $x \rightarrow \frac{x}{a}$, where $a$ is the period of the potential, and the dimensionless time $\tau = t \frac{k_BT}{\gamma a^2}$ and we will model the periodic part \eqref{ten4} by the sinusoid $V_0(x) = u_0 sin^2(2 \pi x)$.

The dependence $\sim r^3$ appears in \eqref{ten4}) due to the fact that such is the dependence of the optical potential in the Rayleigh regime \cite{Bulgakov2019}. In the second term, which leads to the slope of the potential, the dependence on the radius is also included, since in our case the slope of the potential will be modeled using the motion of the fluid with a constant velocity, and the friction force in this case is proportional to $r$ (Stokes formula). The characteristics of the discrete model $K_{\pm}$, $v$, $D$ are obtained depending on the radius $r$. We also introduce the function $J = \frac{1}{v} \frac{d v}{d r}$, then $Jdr = d v/v$~--- relative change in the particle velocity. 

Let at the moment $\tau = 0$ only one well $p_1(\tau = 0) = 1$ is populated by particles. Then the distribution function evolves in two ways. The distribution function as a whole will drift with an average velocity $v(r)$ and simultaneously spread out in width with a time-dependent variance $\sigma = \sqrt{2D\tau}$. Particles of different sizes move differently. Let us carry out a simple estimate of the event when the probability distribution function $p_j(\tau)$ for close sizes $r_1 < r_2$ is spatially separated. In the other words, particles of sizes $r_1$ and $r_2$ are located in different potential wells. The distribution function is Gaussian, the maximum of which is at the position $x_r = v(r) \tau $ and its variance $ \sigma(r) = \sqrt{2D(r) \tau}$, therefore the distribution functions for sizes $r_1$ and $r_2$ will practically cease to overlap if the following condition for the time $\tau_c$ is satisfied
\begin{equation}\label{ten5}
(v(r_1) - v(r_2)) \tau_c \gtrsim 3 (\sqrt{2 \tau_c D(r_1)} + \sqrt{2 \tau_c D(r_2)})             \mbox{ .}
\end{equation}
The expression \eqref{ten5} is simplified if we take into account that $D(r_1) > D(r_2)$, and $(v(r_1) - v(r_2))/v(r_1) \approx -J(r_1) \Delta r $. 
Then we get
\begin{equation}\label{ten6}
v(r_1) \tau_c \gtrsim  \bigg( \frac{2 D(r_1)}{v(r_1)}  \bigg) \bigg( \frac{v}{\Delta v}    \bigg) ^2       \mbox{ .}
\end{equation}

\begin{figure}[h!]
\centering
\includegraphics[width=8.5cm]{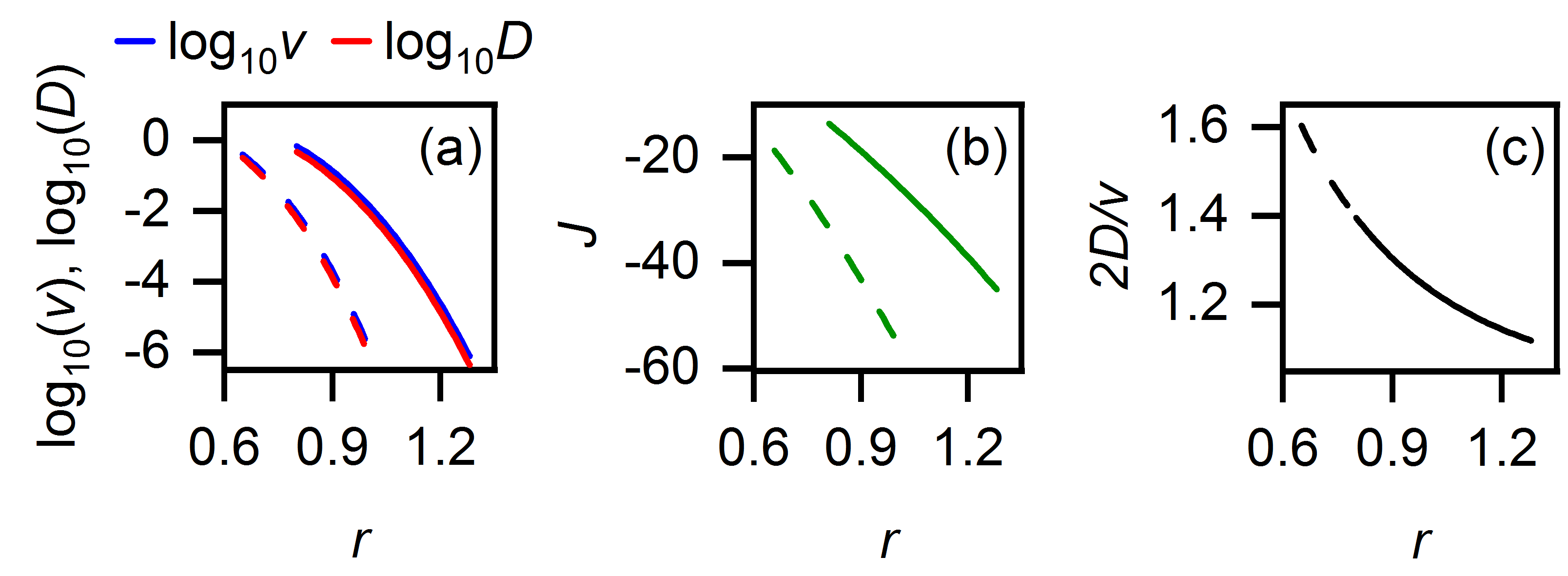}
	\caption{\label{ris:fig4} 
 Calculated drift velocity $v(r)$, the diffusion coefficient $D(r)$, the functions $J(r)$, $2D(r)/v(r)$ in the case of potential \eqref{ten4} for the following values $u_0=10$, $e_0=4.5$~--- solid lines, $u_0=20$, $e_0=4.5$~--- dash lines.}
\end{figure}

Let us take for definiteness $\Delta v/v = 1/2$ (particles move with velocities that differ by a factor of two as in numeric experiment).
In this case we have
\begin{equation}\label{ten7}
v(r_1) \tau_c \gtrsim  144 \bigg( \frac{2 D(r_1)}{v(r_1)}  \bigg);\mbox{  }  \Delta r \gtrsim - \frac{1}{2 J}       \mbox{ .}
\end{equation}
Such simple formula makes it possible to estimate the sorting sensitivity $\Delta r$ (the difference in particle sizes that can still be separated) and the spatial and temporal scales required for this.
Figure \ref{ris:fig4} shows numerical data for the potential \eqref{ten4}. The drift velocity and diffusion coefficient decrease strongly with changing radius $r$, but the ratio $2D/v$ changes only slightly.

An important characteristic of the sorting is the function $J(r)$, which gives an estimate of the sorting sensitivity. If we follow the formula \eqref{ten7}, $ \Delta r \gtrsim 0.02 $, $v \tau \gtrsim 173$, $J=-25$, for $r = 1$. From these data we can conclude that particles of radius $r_1 = 0.98$, $r_2 = 1$ will be spatially separated on a scale of about 200 potential wells in a time $\tau \sim 1.2 \cdot 10^4$. Simple estimates based on (\ref{ten7}) describe the real time dynamics quite well. 

\begin{figure}[h!]
\centering
\includegraphics[width=8.5cm]{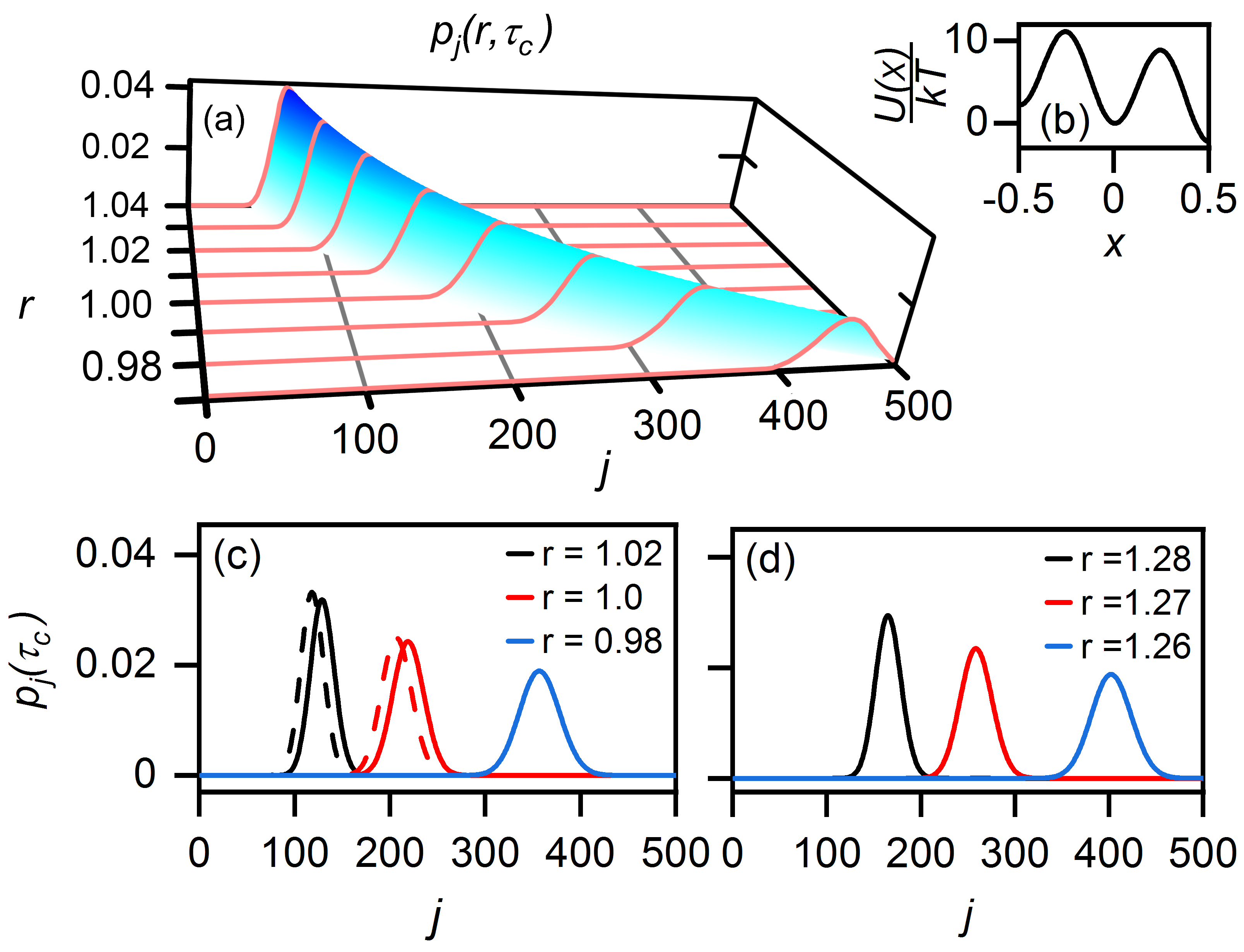}
	\caption{\label{ris:fig5} Separation particles differing in size during diffusion. (a) Dependence of the probability $p_j(\tau_c)$ of filling $j$-th well after $\tau_c=1.5 \cdot 10^4$  in an tilted potential (b) with parameters $u_0=10$, $e_0=4.5$. (c) The same as (a), but only for three types of particles $r=0.98,1.0,1.02$. (d) The same as (a), but for particles of size $r=1.26,1.27,1.28$ after $\tau_c=2.1 \cdot 10^8$. Dashed line (c) shows the distribution function for the sizes  $r=0.99,1.0,1.01$, but for parameters $u_0=20$, $e_0=4.5$ after $\tau_c=1.6 \cdot 10^8$.}
\end{figure}

Figure \ref{ris:fig5} shows the results of the numerical solution of equation \eqref{ten} in the case of potential \eqref{ten4}. Particles of different sizes $r_1 = 0.98$, $r_2 = 1$, $r_3 = 1.02$ are placed in a potential well with index $j=1$ at the initial moment of time. 
After a time $\tau = 1.5 \cdot 10^4$ the mixture of three types is spatially separated and the entire separation process required the participation of no more than 450 potential wells.

The function $J(r)$ decreases monotonically and this means that the sensitivity increases with increasing particle radius. For example, for $r = 1.27$, $-1/2 J \approx 0.01$ and therefore we can expect that mixture of two types of particles $r_1 = 1.27$, $r_2 = 1.28$ can again be separated by position at a distance of several hundred wells, which is fully confirmed by numerical calculations of the distribution function, as shown in Fig. \ref{ris:fig5}(d). The draw-back is that the sorting time has increased by $10^4$ times compared to the previous example, since the drift velocity of the particles has also decreased by the same factor. Fine sorting requires deeper potential wells, respectively, the diffusion process is greatly slowed down. In the last example, the physical sorting time even for nanometer particles will be several days and this is already at the limit of acceptability. Sorting time can be reduced by increasing the slope of the potential, but the effect is not very significant, at best such optimization can reduce sorting time only by a few times and not by orders of magnitude.

It is also possible to increase the sorting sensitivity in the vicinity of $r = 1$, for this we need to increase the potential $u_0$. In the figure \ref{ris:fig5}(c) the case $u_0=20$ (dashed lines) is considered, when $u_0$ increased by 2 times. This gave us the possibility of sorting three types of particles $r = 0.99, 1, 1.01$ without changing the number of wells involved in the process, while the sorting time again increased sharply.

The general conclusion from the analyzed examples is that the use of optical tilted potential opens new the possibilities of  sorting particles by radius over a sufficiently small spatial interval due to Brownian motion. This occurs because of the dependence of the optical potential on the particle radius which in turn  greatly changes the diffusion process in the deep well regime.

For sorting, it is first necessary that there are an average drift velocity of particles through the potential. Clearly, this is ensured by the nonequivalence of the wells, which breaks the symmetry of the system (in our case, due to the motion of the fluid).

\begin{figure}[h!]
\centering
\includegraphics[width=8.5cm]{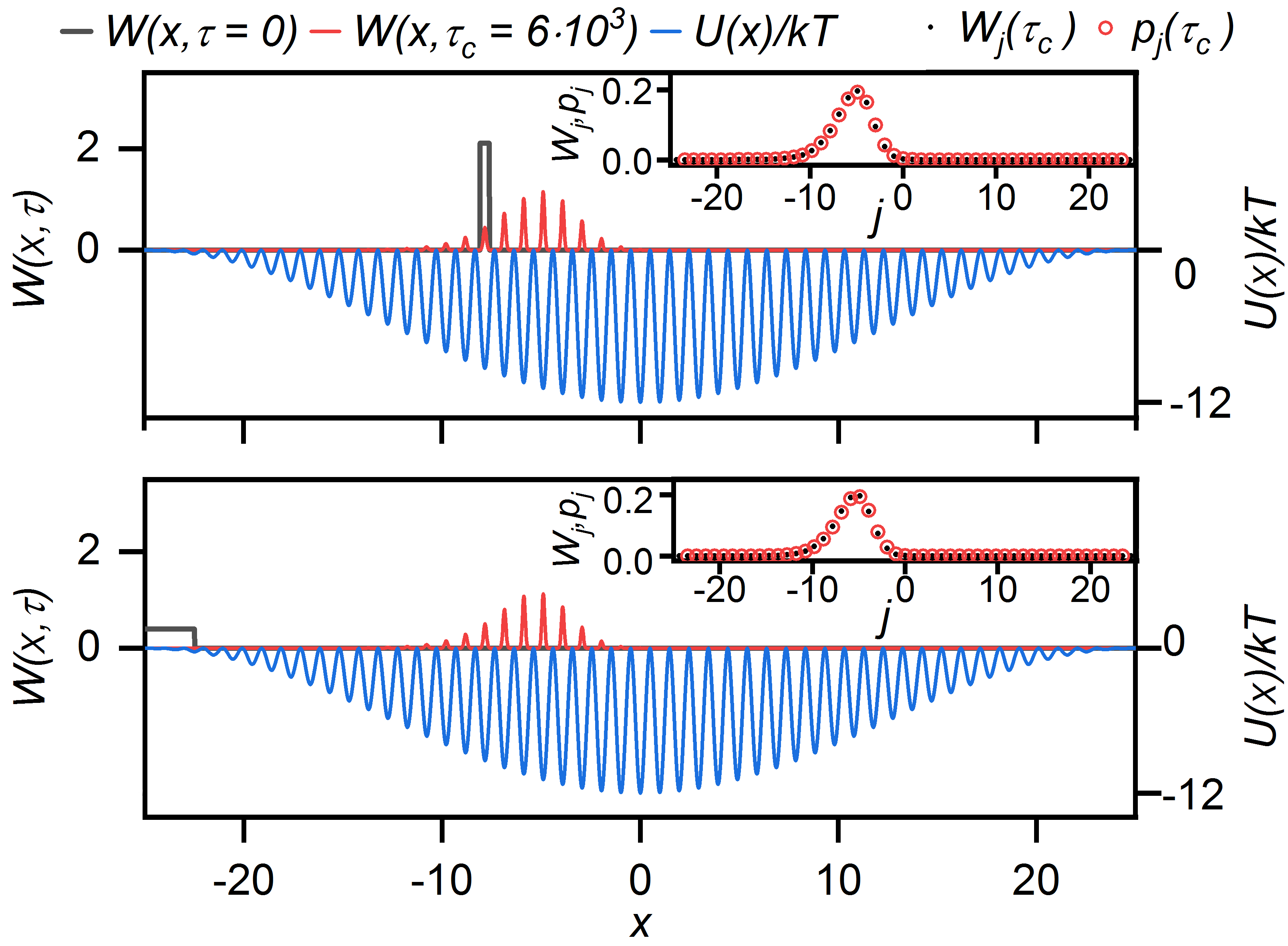}
	\caption{\label{ris:fig6} Diffusion in a multiwell potential \eqref{ten8} $N=51$, $u_0=12$, $r=1$ --- blue line. Black line~--- distribution function at the initial time $\tau=0$. Red line~--- distribution function at the time $\tau_c=6 \cdot 10^3$. (inset) Black dots~--- probability of finding a particle in the $j$-th well $W_j$, red rings~--- $p_j$ probability calculated using the discrete model. (a) Initial distribution inside the potential, (b) initial distribution outside the potential.}
\end{figure}

This is not the unique way of sorting, now we will consider another idea that does not use fluid motion. It is well known that in finite dielectric gratings the wave function of the excited resonant mode becomes quasi-periodic with an envelope in the form of a standing wave \cite{Bulgakov2023,Bulgakov2023b,Bulgakov2019}. Then for the optical potential $\sim |E|^2$ the depth of the potential wells increases from the periphery to the center. A multi-well profile close to the real one can be modeled by a single function.
\begin{equation*}
\frac{U(x)}{kT} = -U_0 \Bigg( \sin \bigg(  \frac{\pi(x-D/2)}{D} \bigg)
\cdot \sin \bigg( \frac{\pi N  (x-D/2)}{D} \bigg) \Bigg) ^2          \mbox{ ,}
\end{equation*}
\begin{equation}\label{ten8}
D=N-1, N=odd \mbox{ .}
\end{equation}
The integer $N$ specifies the number of wells. The potential is shown in the figure \ref{ris:fig6} by blue line for $N=51$. The potential on the left and right slopes of the envelope changes monotonically. This affects both the position of the minima and the depth of the wells. 
On the slopes we can consider it to be tilted, in which the depth of the wells changes simultaneously. In this case, two matrix elements in the equation \eqref{ten} are not enough, and it is necessary to calculate the transition probabilities $K_{j,j \pm 1}$ for each individual well \cite{Nguyen2016}. Let us check the applicability of the discrete model \eqref{ten} for the nonperiodic potential in the form \eqref{ten8}. At the initial time $t=0$ one of the wells with an initial constant probability density (black line) is populated, after a sufficiently long time the particle populates several neighboring wells with a probability density in the form of a comb of peaks (red line). The black dots in the figure are the probability of populating each well $W_j(\tau)$, and the red rings are the probability $p_j(\tau)$ according to the discrete model. As follows from the figure, the agreement between the direct solution of the FP equation and the approximation using \eqref{ten} is very good. Even if at the initial moment the outer wells are populated, as in the figure \ref{ris:fig6}(b), where the applicability of the discrete model is questionable, the agreement nevertheless looks quite satisfactory if the diffusion time interval is chosen to be long. The reason is that the wave packet is quickly drawn into the potential towards the center, but there the discrete model is already justified, and then the diffusion process slows down sharply.

\begin{figure}[h!]
\centering
\includegraphics[width=8.5cm]{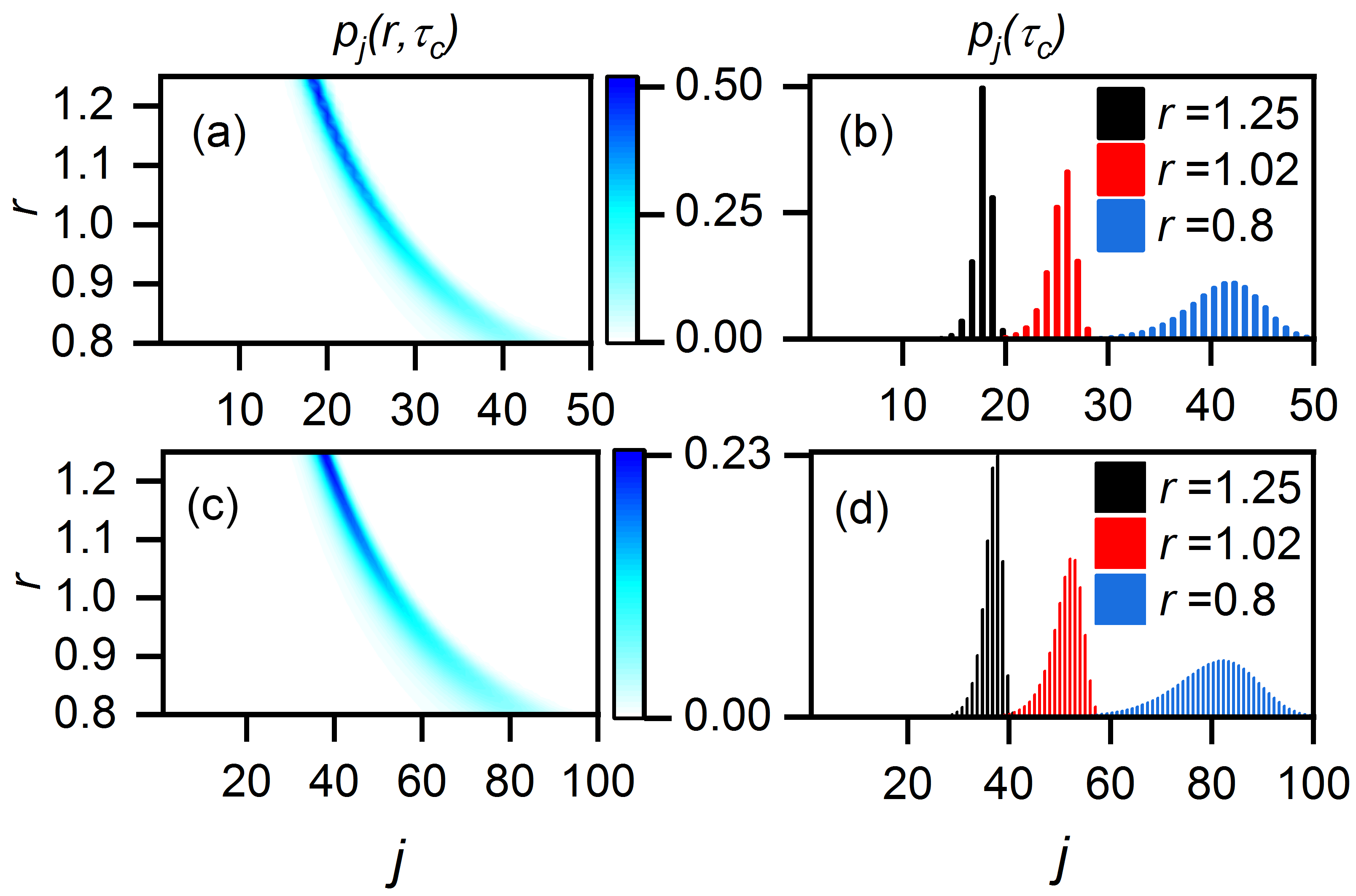}
	\caption{\label{ris:fig7}  Distribution function $p_j(r,\tau_c)$ for different particle sizes (a) $u_0=30$, $\tau_c=5 \cdot 10^5$, $N=101$, (c) $u_0=30$, $\tau_c=2.5 \cdot 10^5$, $N=201$, (b) and (d) for three types of particles  $r=0.8, 1.02, 1.25$. 		
  Only the occupancy of half of the wells is shown from 1 to $N/2$, where the probability may differ from zero.}
\end{figure}

Now let us consider the possibility of sorting particles by introducing into the amplitude of the potential $U_0$ \eqref{ten8} a dependence on the radius in the form $U_0 = u_0 r^3 $.As in the previous case, the sorting strongly depends on both the parameter $u_0$, which controls the potential depth and on the number of wells $N$. Figure \ref{ris:fig7} shows numerical calculations of the distribution function $p_j(\tau_c)$ at a certain fixed time $\tau_c$ for different number of wells $N=101, 201$. The distribution function sharply depends on the radius $r$ and is well localized in space. For this choice of $u_0$ and $\tau_c$, particles with radii $r = 0.8, 1.02, 1.25$ turned out to be spatially separated.

The distribution functions in the figure \ref{ris:fig7} look similar, but include different numbers of wells. From the point of view of the obtained result, there is no particular difference whether 101 or 201 potential wells are used for sorting. However, in practice, it is possible to obtain an optical potential by exciting a resonant mode with a sufficiently large gain only when the number of periods of the structure is $\sim 100$ or more. In the next section, we will turn to the practical case of exciting a tilted potential with a large number of wells, using a PhC dielectric structure. We will show that our results obtained within the framework of simple models are completely reproduced in realistic conditions for sorting nanosized particles.

\section{Sorting of nanoparticles by a two-slabs structure upon excitation of the BIC mode}

A periodic tilted quasi-one-dimensional potential for a Rayleigh particle can arise upon excitation of a high-Q optical mode in the PhC structure shown in Fig.\ref{ris:fig1}.
\begin{equation}\label{ten9}
U(x,y) = -  \frac{\varepsilon-\varepsilon_s}{\varepsilon+2\varepsilon_s} \pi r^3 |\bf{E}(x,y)|^2           \mbox{ ,}
\end{equation}
here $\varepsilon=3$, $\varepsilon$ is the permittivity of a particle of radius $r$, and $\varepsilon_s=1.33^2$ is the permittivity of the surrounding liquid (water). At normal incidence of a plane TM polarized wave, a resonant TM mode will also be excited, for which the electric field is directed along the $z$ axis and does not depend on the $z$ coordinate. For the purpose of sorting nanoparticles, it is desirable to fulfill two conditions: 1) the antinodes of the periodic potential are between the slabs, 2) the resonant mode must have a high Q-factor, so that upon excitation the electric field is significantly enhanced between the slabs. Ideal modes with a very high Q-factor (formally infinite) are BICs. Of course, in a real situation BICs will have a finite $Q$ factor due to structural fluctuations, non-radiative losses, the finiteness of the PhC sizes, etc. 
\begin{figure}[h!]
\centering
\includegraphics[width=8.5cm]{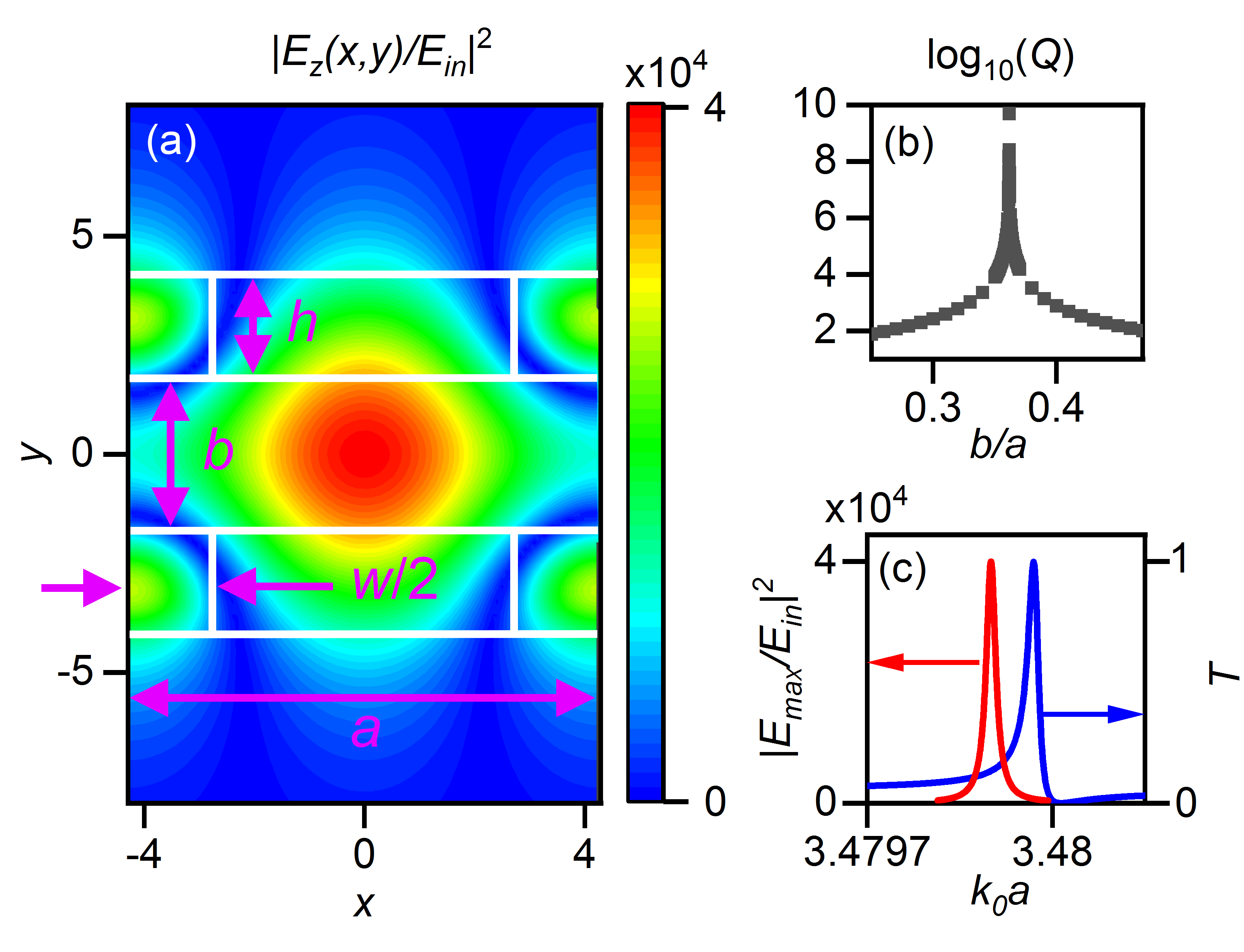}
	\caption{\label{ris:fig8} 
(a) Electric field $|E_z(x,y)|^2$ for a accidental BIC at a wavelength in vacuum $\lambda = 1.55~\mu$m. Geometrical dimensions of the double lattice: $a=0.8585~\mu$m, $w/a=0.3$, $h/a=0.25$, $b/a=0.356$. 
(b) Dependence of the quality factor of the resonant mode $Q$ on the distance $b$ between the slabs. 
(c)(left) EM field gain in the structure $\big|E_{max}/E_{in} \big|^2$ as a function of wave vector $k_0a$, $E_{max}$ is the maximum value of the field in the structure $E_z(x,y)$, $E_{in}$ is the amplitude of the incident plane wave.
(c)(right) Transmission coefficient of a plane wave at normal incidence when $b$ is set to $Q=10^5$ as a function of $k_0a$. }
\end{figure}

After a certain optimization, we settled on the BIC at the $\Gamma$-point, which is commonly called accidental \cite{Koshelev2019} since its appearance requires adjusting the geometric parameters of the structure, in our case this is the distance between the slabs (Fig.\ref{ris:fig8}(a)). At a certain critical value of the distance $b_c$, the coupling of the resonant mode with the radiative continuum disappears, the mode becomes non-radiative and localized along the $y$ direction. The localized mode cannot be excited by an external source, so it is necessary to deviate a little from the value $b=b_c$, while we will always have a large, but finite $Q$ factor. As follows from Fig.\ref{ris:fig8}(a), the mode we selected also has a pronounced wells relief between the PhC slabs and is practically indistinguishable from the BIC.

\subsection{Sorting by liquid movement}

\begin{figure}[h!]
\centering
\includegraphics[width=8.5cm]{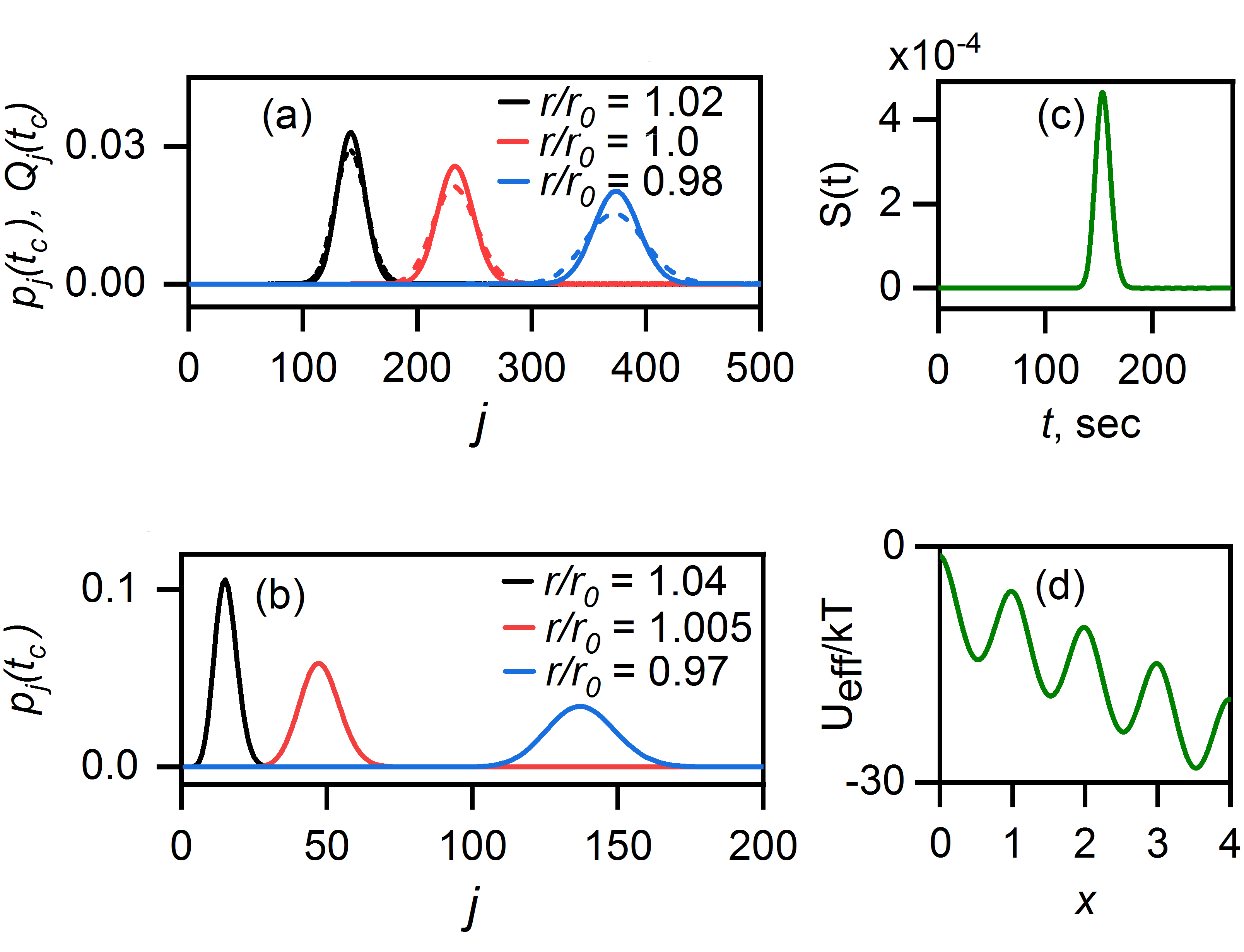}
	\caption{\label{ris:fig9} 
 (a) Distribution function $p(j)$ for three particle sizes, $r_0=3$~nm (a) in case $W=9.4 \big[$mW$/\mu$m$^2\big]$, $v_{f,x}= 430 \big[\mu$m$/$s$\big]$, $r/r_0=0.98,1.0,1.02$, at a point in time $t_c=155$~s. The dashed line shows function $Q_j$ with the same parameters as solid line, but in the case of finite length z: $l=3434~\mu$m (length by $z$), $v_{f,z}=22.5 \big[\mu$m$/$s$\big]$. (c) Probability flow through
the boundary $S(t)$ at $l=3434~\mu$m and $v_{f,z}=22.5 \big[\mu$m$/$s$\big]$. (b) Distribution function $p(j)$ in case $W=12.8 \big[$mW$/\mu$m$^2\big]$, $v_{f,x}= 950 \big[\mu$m$/$s$\big]$, $r/r_0=0.97,1.005,1.04$, at time $t_c=90$~s. (d) Calculated effective potential in
which diffusion occurs.}
\end{figure}

Let us fix the distance $b/a=0.356$ between the slabs so that the $Q$ factor of the resonant mode is large enough, say $10^5$. With this choice, we have a significant enhancement of the near-field (see Fig.\ref{ris:fig8}(c)) when excited by a plane TM wave. The tilt of the periodic potential ensures the movement of the liquid between the slabs with a constant velocity $v_{f,x}$ along the $x$ axis. For practical calculations, it is best to use dimensionless units, then for the chosen mode (Fig.\ref{ris:fig8}(a)) we obtain the following potential for resonant mode exitation
\begin{multline}\label{twenty}
\frac{U(x',y')}{kT} = \frac{\varepsilon-\varepsilon_s}{\varepsilon+2\varepsilon_s} 18.67 \cdot 10^7 \cdot W \Bigg[\frac{mW}{\mu m^2} \Bigg] \cdot\\
\cdot U_{norm}(x',y')   
r'^3 - 3v_{f,x} \Bigg[\frac{\mu m}{s} \Bigg]  r'x'               \mbox{ ,}
\end{multline}
here $r'=r/a$ dimensionless radius of particle, $(x',y')=(x,y)/a$, $a$ is the lattice period, $W$ is the incident plane wave power measured in $[$mW$/\mu$m$^2]$, $v_{f,x}$ is the fluid velocity measured in $[\mu$m$/$s$]$, $U_{norm}$ is the potential profile normalized so that $\max |U_{norm}(x',y')|=1$. The FP equation for the probability density $P(x',y',\tau)$ takes the form
\begin{equation}\label{twenty_1}
\frac{\partial P(x',y',\tau)}{\partial \tau} = 
\sum_{j=1,2} \frac{\partial}{\partial x'_j} \Bigg( \frac{\partial}{\partial x'_j} + 
\frac{\partial \Big( \frac{U(x',y')}{kT} \Big)}{\partial x'_j} \Bigg) P(x',y',\tau)          \mbox{ ,}
\end{equation}
where $\tau= 0.3856t/r'$ is the dimensionless time, and the temperature is assumed to be $T=300$~K. In particular, formula \eqref{twenty} shows that for a particle with radius $r=1$~nm and $W \sim 1$~mW$\mu$m$^2$, $max|(U(x',y'))|/kT \sim 1$, in other words, for particles of the order of a large molecule in size, the optical potential still competes with Brownian forces. The solution of the FP equation (\ref{twenty_1}) with the real optical potential taking into account the fluid motion (\ref{twenty}) was found in the manner described in Section I by going to the one-dimensional PhC equation with the potential $U_{eff}$, and then to the discrete model. The results of calculating the discrete well distribution functions are shown in Fig.~\ref{ris:fig9}(a,b) for sorted particles with sizes in the vicinity of $r_0=3$~nm. The figure captions indicate the power of the incident plane wave and the velocity of the fluid between the slabs along the $x$ axis, and Fig.~\ref{ris:fig9}(d) shows the calculated effective potential in which diffusion occurs. In the first example, using $\sim 500$ potential wells (the rest are not populated by particles) in a time of $t \sim 150$ seconds, it is possible to spatially separate a mixture of three types of particles with sizes $r/r_0 = $ 0.98, 1, 1.02 . In the second example, only $\sim 200$ wells are used for sorting, but in this case the separation by size is coarser $r/r_0 = $ 0.97, 1.005, 1.04. 

After the spatial separation of different types of particles has occurred, it is necessary to propose a method for removing particles from the operating area (where the optical potential acts). Below we will discuss one of the options, which seems to be quite practical. First, we will assume that the size of our PhC structure along the $z$ axis is finite, although quite large. Second, the fluid moving between the slabs has a velocity component along the $z$ axis as well. The idea is that particles of different sizes penetrate into the region between the slabs through a small hole (Fig.~\ref{ris:fig1}), and then diffuse in the moving flow, both along the $x$ and $z$ axis. During diffusion along the $x$ direction close sizes differing by a few percent are spatially separated after a long time interval due to the action of optical forces. While diffusion along the $z$ direction occurs without any participation of optical forces, and therefore, with such a small difference in size, looks almost identical for different particles. This suggests that it is necessary to choose the length of the PhC along the $z$ axis so that at the moment when particles of different types have already separated spatially along $x$, they all together approach the boundary of the structure in the $z$ direction and almost simultaneously leave the operating area. After the particles leave the operating area, each type gets into its own limited cell of space (Fig.~\ref{ris:fig1}) and for this reason cannot mix further. At this point, the sorting is complete. Let us carry out an analytical calculation substantiating this idea. The distribution function over the wells taking into account the dependence on $z$ has the form
\begin{equation}\label{twenty_2}
P_n(z,t) = p_n(t) W_z(z,t)          \mbox{ ,}
\end{equation}
here $W_z(z,t)$ is the probability density along the coordinate $z$. Since the variables are separated, we obtain a separate equation for the function $W_z(z,t)$
\begin{equation}\label{twenty_3}
\frac{\partial W_z(z,t)}{\partial t} = \frac{\partial }{\partial z} \Bigg( kT \frac{\partial }{\partial z} - F_z \Bigg) W_z(z,t)       \mbox{ ,}
\end{equation}
where $F_z$ is the Stokes force acting in the direction of $z$, $F_z = 6\pi \eta r v_{f,z}$. We impose the simplest absorption boundary conditions $W_z(z=l,t) = 0$ on the boundaries of the PhC structure, then the probability flow through the boundary $z=l$ has the form
\begin{equation}\label{twenty_4}
S(t) = -kT \frac{\partial W_z(z,t)}{\partial z} \bigg|_{z=l}              \mbox{ .}
\end{equation}
The probability of leaving operational region while in the $j$-th well
\begin{equation}\label{twenty_5}
Q_j = \int_0^\infty S(t) p_j(t) dt              \mbox{ .}
\end{equation}
It is necessary to select the parameters $l$ and $v_z$ so that the peak of the probability flow $S(t)$ falls exactly at the moment of time when the spatial separation of particles of different types has already occurred. In this case, as we expect, the probability distributions $Q_j(r)$ will also be spatially separated (will not overlap), as can be seen from Fig.~\ref{ris:fig9}(a) dashed line. At the exit from the operational region $z>l$, the particles fall into separate containers with boundaries (see Fig.~\ref{ris:fig1}) and this is where the sorting is complete. It is clear that such a sorting scheme can work continuously, since the sorted particles never accumulate in the operational region and are removed at the appropriate moment.

\begin{figure*}[ht]
\center\includegraphics[width=18cm]{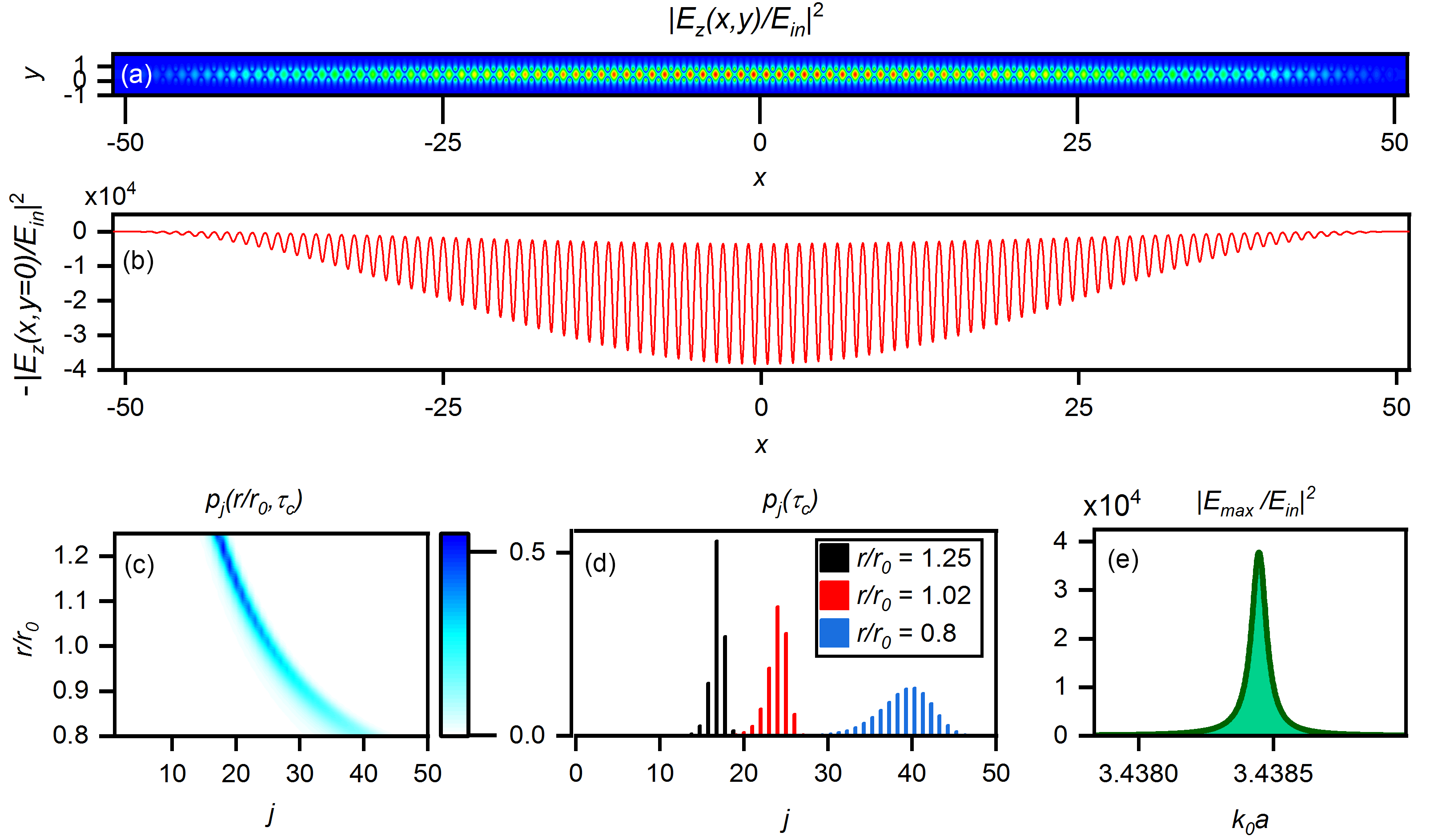}
	\caption{\label{ris:fig10} 
 (a) Profile of the resonance mode $|E_z(x,y)/E_{in}|^2$ excited at normal incidence of a plane wave with Wave vector $k_0a=3.4384$ at which the maximum field gain inside the structure is achieved, the number of potential wells $N=101$,
  (b) profile of $|E_z(x,y=0)/E_{in}|^2$ at the center of the structure, 
 (c) probability distribution $p_j$ for different sizes $r/r_0$, $r_0=3 nm$, after time $t_c=1800 s$, (d) the same as (c) for three types of particles $r/r_0 = 0.8, 1.02, 1.25$. Incident wave power $W=9.1 \big[$mW$/\mu$m$^2\big]$. (e) EM field gain in the structure $\big|E_{max}/E_{in} \big|^2$ as a function of wave vector $k_0a$, $E_{max}$ is the maximum value of the field in the structure $E_z(x,y)$, $E_{in}$ is the amplitude of the incident plane wave.}
\end{figure*}
The fact that fine sorting of particles with sizes close to $r_0=3$nm is chosen as an example in Fig.\ref{ris:fig9} is not of fundamental importance, due to the obvious scaling in the potential (\ref{twenty}). If we make a parameter substitution of the form $\widetilde{W}=W/\xi$, $\widetilde{v}=v/\xi^{\frac{1}{3}}$, $\tilde{r}' = r'\xi^{\frac{1}{3}}$, where $\xi$ is the scaling parameter, then the expression \eqref{twenty} in new parameters will remain unchanged. The constancy of the optical potential means that, say, a decrease in the incident wave power by a factor of 2 ($\xi=2$) and the liquid velocity by $2^{\frac{1}{3}}$ corresponds to an increase in the size $r_0=3$~nm $\cdot 2^{\frac{1}{3}}$ of finely sorted particles in the figure~\ref{ris:fig9}.
Thus, we conclude that by appropriately adjusting the physical parameters $W$, $\textbf{v}_f$ we can finely sort any particle size.

\subsection{Sorting using the optical potential created by the finite PhC}

If the PhC structure consists of a finite number of periods $N$ along the $x$ axis, then the ideal undamped BIC mode transforms into an ordinary mode with a finite lifetime, while the $Q$ factor usually scales according to the power law $Q \sim N ^{\alpha}$ \cite{Bulgakov2023b}. For the accidental BIC under consideration, $\alpha=3$, and for sufficiently large $N$ the $Q$ factor reaches a significant value. In this case, the mode profile in each cell looks the same as in the case of an infinite PhC, but the amplitude decreases toward the edges of the structure (Fig.\ref{ris:fig10}(a,b)). The envelope resembles a standing half-wave. As a result, we obtain a linear chain of potential wells, arranged along the $x$ axis, with a smoothly changing depth (Fig.\ref{ris:fig10}(b)). 

We consider the case of $N=101$ periods, the mode shown in Fig.\ref{ris:fig10}(a,b) has a quality factor $Q=5 \cdot 10^4$. Under resonant excitation by a plane wave incident normally to the surface with amplitude $E_{in}$ inside the PhC the amplitude of the electric field is significantly enhanced (Fig.\ref{ris:fig10}(e)), $\big|E_{max}/E_{in} \big|^2 \sim 4 \cdot 10^4$. Naturally, the optical potential \eqref{ten9} between the slabs is increased the same number of times.

For the optical mode we have chosen, resonant potential has the form
\begin{equation}\label{twenty_6}
\frac{U(x',y')}{kT} = 0.484 \frac{\varepsilon-\varepsilon_s}{\varepsilon+2\varepsilon_s} \cdot r'^3 10^9 |E_{norm}|^2 \cdot W \bigg[\frac{\text{mW}}{\mu \text{m}^2} \bigg]                \mbox{ ,}
\end{equation}
here as before $|E_{norm} (x',y')|$ is the mode profile normalized by the condition $max|E_{norm} (x',y')|=1$. 

For the incident wave power $W=1 [$mW$/\mu$m$^2]$ and $r \sim 1$~nm, we again obtain $max|(U(x,y))/kT| \sim 1$. The estimate shows that even for such a small particle size, the optical forces seriously compete with Brownian ones, and for $W \gg 1 [$mW$/\mu$m$^2]$, strong localization of particles inside the wells occurs and the diffusion process slows down significantly. And now diffusion occurs due to the nonzero probability of particle transition from a well to an adjacent well. This process is still correctly described by the discrete model. The results of calculating the probability distribution function for different sizes $r$ are shown in Fig.\ref{ris:fig10}(c,d).

For sorting only the left part of the optical potential in Fig.\ref{ris:fig10}(b) is used, consisting of 50 potential wells. In this case, three types of particles \ref{ris:fig10}(d)
were sorted during the time $\sim 1800$ seconds, the sizes of which differ by $\sim 20\%$. This result agrees well with the calculations of the model (Fig.\ref{ris:fig7}), when the effective potential was approximated by the expression \eqref{ten8}. The use of a PhC with a large number of elementary cells $N$ significantly enhances the near field, since the $Q$ factor of our quasi-BIC mode increases $\sim N^3$. And this makes it possible to use smaller values of the power $W$ of the exciting wave, however, our calculations like the earlier model calculations (Fig.\ref{ris:fig7}), did not reveal a serious effect of increasing the size $N$ on the sorting sensitivity without significantly increasing sorting time. It can be estimated that only mixtures of particles with sizes different by more than $20\%$ can be ideally separated.

\section{ Conclusion}

The possibility of manipulating nanometer particles is a fundamental technological task of modern science. Conventional optical traps (tweezers) cannot provide sufficient optical force to influence the motion of dielectric particles smaller than 100~nm and even more so to separate mixtures of such particles with high resolution. Currently, various near-field technologies are used for these purposes, enhancing the rigidity of the localizing potential. In this paper, we develop the idea of fine sorting of dielectric particles by their size during Brownian diffusion in a strong periodic optical potential consisting of a sequence of deep wells strongly localizing particles. Such a quasi-one-dimensional potential can be obtained by exciting a resonant mode with a high $Q$ factor in a PhC structure of two parallel slabs (quasi-BIC mode). Diffusion in such a potential occurs as a result of the Kramer mechanism of transition through the barrier. Due to this, the diffusion becomes very sensitive to the particle size. Two methods of sorting are proposed for the spatial separation of particles of different sizes. In the first method it is assumed that the PhC is infinite along the diffusion axis, and the liquid moves between the slabs to induce an effective tilt of the optical potential.
In our calculations on the scale of only several hundred potential wells, it was possible to separate three types of particles whose sizes lie in the vicinity of 3~nm and differ by only $2\%$. In the second method, liquid motion is not used. The potential profile required for sorting is achieved by resonant excitation of the quasi-BIC mode that is finite along the PhC diffusion axis. In this case, particles differing in size by more than $20\%$ can be spatially separated precisely using only 100 potential wells. In both cases, the incident wave power required was $\sim 10 [$mW$/\mu$m$^2]$. A method for removing sorted particles from the operating region is also proposed and discussed to ensure the continuity of the sorting process.

\begin{center}
\bf{ACKNOWLEDGEMENTS}
\end{center}

The work was supported by Russian Science Foundation Grant No. 22-12-00070, \url{https://rscf.ru/project/22-12-00070/}.

\bibliographystyle{apsrev4-1}
\bibliography{ref}

\end{document}